\pdfoutput=1
\documentclass[twocolumn,times,tighten,floatfix]{aastex62}
\usepackage{natbib}

\usepackage{epsfig}
\usepackage{amsmath,amssymb}
\usepackage{ulem}
\usepackage{graphicx}
\usepackage{upgreek}
\usepackage{float}
\usepackage{comment}
\usepackage{placeins}
\usepackage[T1]{fontenc}
\usepackage[utf8]{inputenc}

\begin{document}

\renewcommand*{\sectionautorefname}{\S\!}
\renewcommand*{\subsectionautorefname}{\S\!}
\renewcommand*{\subsubsectionautorefname}{\S\!}

\font\sevenrm=cmr7 scaled 1000

\newcommand{\Msun}{\ifmmode M_\odot \else $M_\odot$\fi}
\newcommand{\Mbh}{\ifmmode M_\text{BH} \else $M_\text{BH}$\fi}
\newcommand{\kms}{\ifmmode {\text{km~s}^{-1}} \else km~s$^{-1}$\fi}

\newcommand{\Hb}{\ifmmode {\rm H }\upbeta \else H$\upbeta$ \fi}
\newcommand{\Hgamma}{\ifmmode {\rm H}\upgamma \else H$\upgamma$\fi}
\newcommand{\dg}{\ifmmode {\rm ^{\circle}} \else ^{\circle}\fi}

\title{\uppercase{The direction of rotation of supermassive black holes is unrelated to the direction of rotation of the host galaxy}}

\author{Loren Gigi}
\affiliation{Department of Astronomy and Astrophysics, University of
California, Santa Cruz, CA 95064, USA}

\author[0000-0003-4888-2009]{C.~Martin Gaskell}
\affiliation{Department of Astronomy and Astrophysics, University of
California, Santa Cruz, CA 95064, USA}

\keywords{galaxies: active -- galaxies: nuclei --  quasars: supermassive black holes -- galaxies: jets -- galaxies: structure}
\begin{abstract}

We compare the apparent directions of rotation in the plane of the sky of active galactic nuclei (AGNs) and their host galaxies. The direction of rotation of the galaxy was inferred from the direction of the spiral arms, while the direction of rotation of the AGN was inferred from spectropolarimetry, where the change in relative polarization position angle (PA) across broad lines is believed to be caused by equatorial scattering. The numbers of co-rotating and counter-rotating AGNs are equal. Studies of the relative position angles of radio jets have implied that there is a ``zone of avoidance'' where jets avoid being in the plane of disk galaxies. We point out that bi-conical narrow-line-region outflows also avoid the plane of the host galaxy. The equal numbers of co-rotating and counter-rotating AGNs exclude the hypothesis that the ``zone of avoidance'' is due to a lack of large tilts of the black hole rotating axis relative to the host galaxy rotation axis. Our results imply that the relative orientations of spin axes are random, at least for the black hole mass range we consider. We propose that changes in the broad-line polarization PA with wavelength that do not closely follow the predictions of the simple equatorial scattering model are a consequence of the scattering dust being clumpy. We note a couple of cases of possible changes in PA over several years, which, if real, could be due to motions of the dust clumps or changing anisotropy of the continuum emission.
\end{abstract}

\section{Introduction}

The source of material feeding supermassive black holes (SMBHs) in active galactic nuclei (AGNs) in spiral galaxies is widely assumed to predominantly be gas from the galactic disk. The gas settles into a disk around the black hole and the Bardeen-Petterson effect \citep{Bardeen+Petterson75} causes alignment of the angular-momentum vectors of the disk and the spin of the black hole through frame-dragging.  The alignment is relatively rapid \citep{Natarajan+Pringle98}. Because, on average, the accreting gas shares the angular momentum of the host galaxy, it was long assumed (for example, by \citealt{Keel80}) that the angular momentum vectors of AGNs and their host galaxies would tend to be aligned. This pre-1981 paradigm treated the position angle (PA) of the minor axes of the isophotes of galaxies as proxies for AGN orientation. 

However, \citet{Ulvestad+81} showed that radio jets in disk galaxies did {\em not} preferentially align with the minor axes of the host galaxies. This result, which was surprising at the time, shifted thinking toward a new paradigm in which AGN orientations are essentially random relative to their hosts. More recently, studies of bi-conical outflows of the narrow-line region (NLR) of nearby active galaxies show that these outflows are also not aligned with the minor axes of the host galaxies (see \citealt{Fischer+13,Fischer+14}). 

As \citet{Hopkins+12} emphasize, the orientation and spin of the black hole in an AGN has wide-ranging consequences. Spin direction and magnitude affect the efficiency of jet launching and the nature of AGN feedback, which in turn regulates star formation, quenches cooling flows, and thus shapes galaxy evolution. Also, spins determine whether gravitational-wave recoils from black hole mergers can eject SMBHs from galaxies. Thus, determining how AGN spin axes relate to their host galaxies is of fundamental importance. 

Gas falling in from large scales tends to spin SMBHs up to near-maximum values (black hole spin parameter $a \approx 0.998$) and align them with the inflowing disk of gas. If, however, accretion is chaotic, angular momentum is randomized and spins remain low even in large accretion events.  \citet{Dubois+14B} used a semi-analytical approach to study spin evolution as a function of cosmic time. They find that sustained cold-gas accretion at high redshift generally aligns black hole spins with the angular momentum of the surrounding gas and increases their magnitude. More recently, at low redshift, as black holes have grown more massive and their host galaxies have become increasingly gas poor, black hole mergers contribute more to black hole growth and alter spin directions. Other recent theoretical studies have begun to include black hole spin explicitly in models of black hole--galaxy co-evolution. For example, \citet{Bustamante+Springel19} introduced a prescription for black hole spin evolution in cosmological simulations, accounting for coherent gas accretion, chaotic accretion, black hole mergers, and allowing AGN feedback efficiencies to depend on spin. Their models predict that massive black holes tend to reach high spins through aligned accretion, while lower-mass systems experience more chaotic growth and lower spins.  They find that massive black holes tend to spin up through coherent, aligned gas accretion, while lower-mass black holes experience more chaotic accretion and mergers, resulting in lower spins.

\citet{Chen+25} have recently examined the relationship between black hole spin, black hole mass, and host-galaxy star formation. Black hole spin estimates were taken from X-ray measurements. Although reservations have been expressed about the reliability of X-ray spin estimates (for an extensive review see Section 3 of \citealt{Reynolds21}), the \citet{Chen+25} study suggests different black hole growth modes at low and high masses.  

A number of studies have sought to quantify AGN/host alignments. \citet{Schmitt+97} noted a ``zone of avoidance'' in the distribution of radio jet position angles, with jets rarely being found close to host major axes -- i.e., a deficit of jets close to host minor axes (but see also \citealt{Kinney+00}). The exact form inferred for the zone of avoidance depends on sample selection and methodology. More recently, \citet{Zheng+24} looked at thousands of galaxies and showed that jets favor directions nearer minor than major axes. For elliptical galaxies, \citet{Schmitt+02} found that the minor axes of nuclear dust disks in elliptical galaxies were also not aligned with radio jets. Their statistical analysis implied that jets are homogeneously distributed over a large region, extending over polar caps of $\theta > 55^\circ$ or $ 77^\circ$, but avoiding lying close to the plane of the dust disks. 

These results imply that radio jets avoid both the plane of the host galaxy if the host is a disk galaxy, and the plane of the dust disk in ellipticals.  An important question is whether these avoidances imply that the angular momentum vectors of the AGNs are generally not being tipped over by large angles relative to the vectors of the stellar or gaseous disks -- whereupon there would be a lack of AGN axes close to the plane of the stellar population or dusty disk -- or whether there is some other explanation. Plausible explanations are jet formation or propagation being inhibited by the denser interstellar medium in the plane of the host galaxy, or the avoidance being the result of some sort of selection effect.

These explanations make distinct, testable predictions for the direction of rotation, which have not previously been examined. If the angular momentum vectors of supermassive black holes are initially parallel to those of the host and then are only weakly perturbed (tilt changes of $\lesssim 60^\circ$, say), then co-rotation should dominate and the zone of avoidance is real. On the other hand, if the AGN angular momentum undergoes large changes such that the relative orientations of the angular momentum vectors are random, then half of the relative angles will be greater than $90^\circ$ (i.e., a complete reversal of direction of rotation). In this case, co-rotation and counter-rotation should occur in similar numbers. Determining directions of rotation of AGNs therefore provides a direct test of these different hypotheses. 

Unfortunately, jet alignment studies alone cannot distinguish between co-rotation and counter-rotation. In this paper, we point out that spectropolarimetry can be used to determine the \textit{direction} of rotation of an AGN and that this can be used to see whether a given AGN is co-rotating or counter-rotating compared with its host galaxy. We explain in Section 3 how spectropolarimetry gives the direction of rotation.  We apply this method in subsequent sections and discuss our results.  In Section 2, however, we first look at whether the orientations for the bi-conical outflows of NLRs also show a ``zone of avoidance'' similar to that of radio jets.

\section{A NLR outflow ``zone of avoidance''?}

\citet{Crenshaw+Kraemer00} showed that spatially-resolved spectroscopy of the NLR is a powerful means of determining many parameters of the biconical outflow of the NLR.  As a result, \citet{Fischer+14} are able to present the misalignment angles, $d\theta$, between the outflows and the rotation axis of the host galaxies.  These are summarized in their Table 1. We show the distribution of these in Fig.~1 compared with the distribution expected for random orientations.  It can be readily seen that, compared with the numbers at small $d\theta$, there is a large deficit of AGNs at large $d\theta$ in the sample of \citet{Fischer+14}. The dashed curve is normalized to the 10 objects in the bottom two bins (least misaligned) and predicts 24 objects in the top two bins with an uncertainty of $\pm 8$ from Poisson statistics.  This compares with the actual number of seven.  The two-tailed significance is $p \sim 0.03$.  The NLR outflows are thus also showing a ``zone of avoidance'' like the radio jets.  

\begin{figure}
\centering
\includegraphics[width=75mm,angle=0]{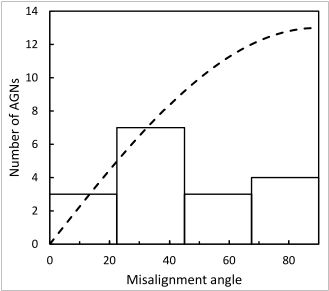}
\caption{The distribution of the angular distances, $d\theta$, between the directions of bi-conical NLR outflows in AGNs to the rotation axes of their host galaxies.  The curve, which is normalized to the number of cases in the lowest angular separation bins, shows the predicted distribution if the directions of the bicones were completely random with respect to the rotation axes of the host galaxies. A deficit of large misalignment angles compared to a random distribution can be seen. Data from \citet{Fischer+14}.}
\end{figure}


\section{Spectropolarimetric determination of direction of rotation of the BLR}


\begin{figure}
\centering
\includegraphics[width=85mm,angle=0]{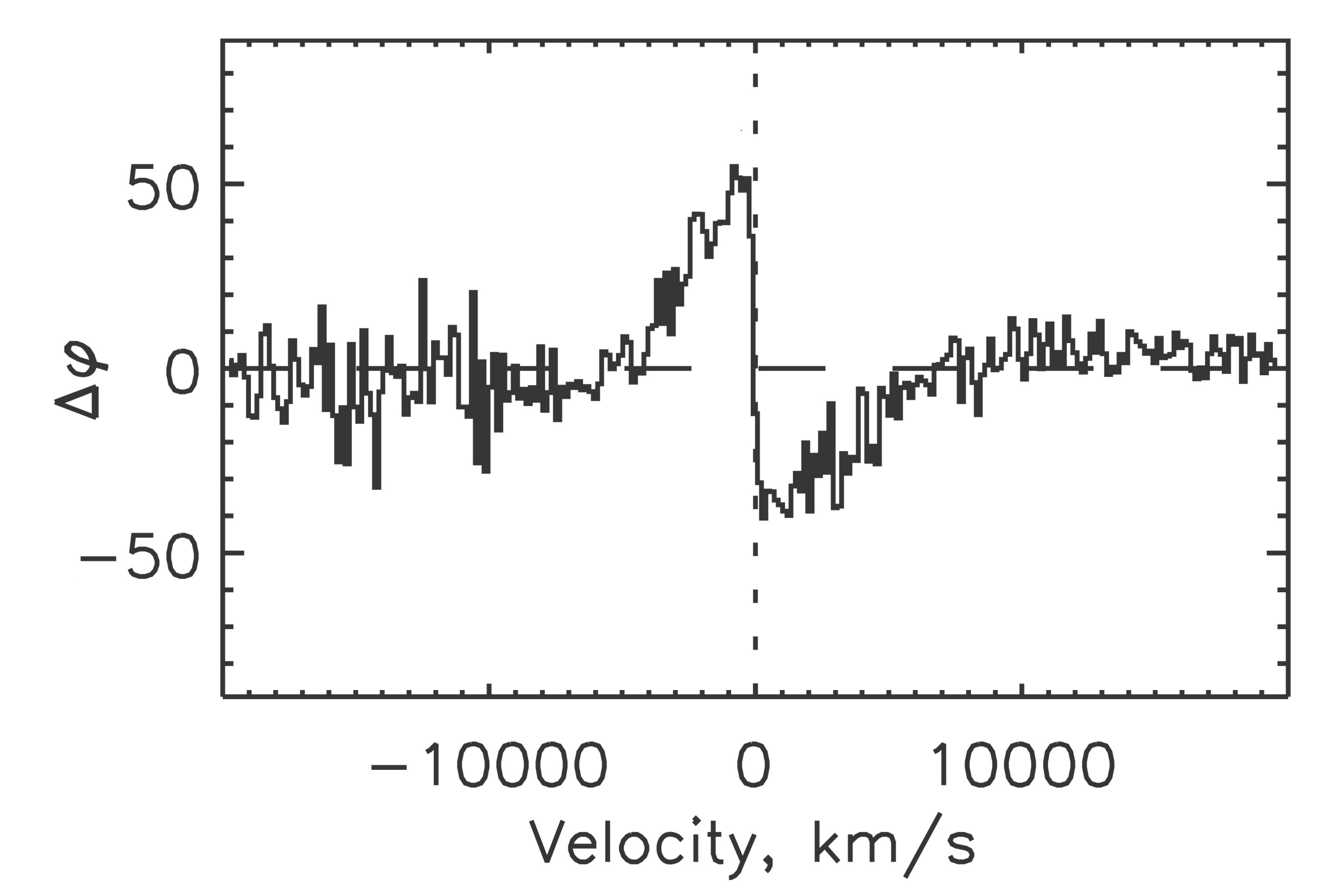}
\caption{The difference, $\Delta\upphi$ (in degrees), in the polarization PA of H$\upalpha$ from the continuum PA as a function of velocity for 3C 273.  (Figure adapted from \citealt{Afanasiev+19})}
\end{figure}

\citet{Goodrich+Miller_94} carried out a spectropolarimeric survey of type-1 AGNs.  Their most striking result was that for AGNs showing internal polarization, all had polarization of the broad Balmer lines, and this line polarization showed a velocity dependence.  Typically, there was a rotation in PA across the line profile.  Goodrich \& Miller interpreted this as the superposition of two components of polarized flux, but \citet{Cohen+99} offered a simpler explanation: the common ``sideways-S'' shape double-humped wavelength dependence with the PA changing in the blue wing of the line and then changing in the opposite direction in the red wing of the line, could be explained by scattering of BLR photons from an equatorial ring (see their Fig.~13 and Figs.~2 and 3 of \citealt{Cohen+Martel02}). This equatorial scattering model was developed further by \citet{Smith+02} and \citet{Smith+05}.  We show a very clear example of the rotation of the PA across a line in Fig. 2. 

The key thing for understanding the geometry of scattering regions producing polarization is that the PA of the {\bf E} vector of a scattered photon is perpendicular to the direction in the plane of the sky of the photon coming towards the scatterer (see, for example, Figs.~2 through 14 of \citealt{Marin+12}). For strict rotational symmetry about our line of sight, there is no net polarization.  However, in most cases, the accretion disk and BLR are inclined to our line of sight.  The resulting foreshortening breaks the rotational symmetry as seen from the earth and hence we see a net polarization (see Fig.~14 of \citealt{Marin+12}).  This explains why most type-1 AGNs show polarization parallel to the axis of symmetry, as was found by \citet{Antonucci83}.

\begin{figure}
\centering
\includegraphics[width=75mm,angle=2]{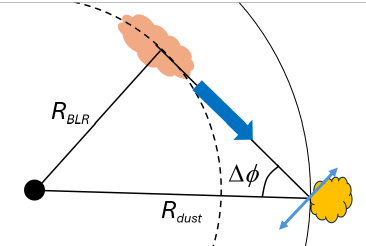}
\caption{Schematic illustration looking from above the accretion disk and BLR to illustrate the scattering geometry producing rotation, $\Delta\phi$, of the PA of polarization of part of a broad-line profile. For simplicity we only consider scattering close to the equatorial plane.  From the viewpoint of the co-orbiting scattering clump of dust shown at $R_{dust}$, the relative radial velocity of the BLR gas shown at a radius $R_{BLR}$ results in a blueshift. Because the BLR gas orbits faster closer to the black hole, the amplitude of the Doppler shift can be seen to decrease with increasing $|\Delta\phi|$. The light blue double-headed arrow indicates the direction perpendicular to the incoming line photon.}
\end{figure}

In Fig.~3 we show how the shifts in the PA of the {\bf E} vector across broad-line profiles arise in the equatorial scattering model. Because the optical-continuum-emitting region of the accretion disk is an order of magnitude smaller than the BLR, and hence can be regarded as a point source, continuum photons travel effectively radially to the dusty scattering region somewhat outside the BLR.  However, as seen from the scattering region, the BLR has a substantial angular extent.  One side is approaching the dust and the other side is receding.  We illustrated the approaching side in Fig.~3 where the BLR is rotating clockwise (CW) as seen from the earth. Photons from the approaching side (the large blue arrow in Fig.~3) are coming from a different direction than photons from the receding side.  This means that the {\bf E} vectors of the blueshifted and redshifted parts of the broad line profiles are approaching the scattering dust different directions. When the BLR is rotating CW as seen from our viewpoint, as in Fig.~3, this results in the PA of the blue wings of the line being rotated clockwise (a decrease in PA) and the red wings of the line (not illustrated) bring rotated counter-clockwise (CCW), i.e., an increase in PA.  For CCW rotation of the BLR, the changes in PA would be the opposite: the PA of the blue wing of the line would increase, while the PA of the red wing would decrease.  If the AGN is exactly face-on, there is no foreshortening of the scattering region, and because there is rotational symmetry about our line of sight, there will be no polarization of the continuum or BLR. In these cases we will not see double-humped profiles in polarized light. Although Smith et al.~depicted the scattering region as a thin ring in the equatorial plane, we note that the scattering does not need to be exactly in the equatorial plane.  To determine the direction of rotation from polarization all is necessary is that the light from the approaching and receding sides of the BLR comes towards the scatterers from different position angles.  Smith et al and \citet{Afanasiev+Popovic15} considered the external scatterers to be a rest, but this is not necessary.  It is only the relative velocity that matters and having the light from the blueshifted and redshifted sides of the broad line approaching the scattered from different PAs.

Infrared reverberation-mapping shows that the hot-dust regions are somewhat larger than the BLR, indicating that the inner dust region lies immediately outside the BLR \citep{Suganuma+06}. Since dust has a much larger scattering cross-section than free electrons, it can efficiently polarize BLR photons. We therefore believe that the scattering region is the dust located just beyond the BLR.  Furthermore, \citet{Gaskell+12} discovered that for NGC~4151 the polarized flux variability lagged behind the variability of non-polarized flux and showed that the scattering region is indeed close to the central regions.

Since the dominant motion of the BLR clouds is rotation (see \citealt{Gaskell09} for a review), if the AGN is viewed close to face-on, the broad lines will appear to be narrower with only the turbulent component of BLR velocity contributing to line broadening (see Fig.~2 of \citealt{Gaskell11} for an illustration).  We thus see narrower broad lines as AGNs are seen more face-on.  However, in polarized flux we are seeing the BLR from the vantage point of the equatorial scattering region, which always gets an edge-on view of the BLR.  As a result, other things being equal, the width of a broad line in polarized light is independent of the inclination of the BLR to our line of sight.  In near face-on AGNs, the width of the line in polarized flux will be greater than what we see in direct (unscattered) light from our viewing position. \citet{Goodrich+Miller_94} found that this was the case for NGC~5548 and Mrk~509. \citet{Martel96} found that the broad emission lines in polarized flux are wider than in total (unpolarized) flux for all of the 13 AGNs where the velocity width of the polarized H$\upalpha$ could be measured. This has been investigated in more detail by \citet{Capetti+21} who find that the ratio of the line width in polarized flux to the line width on total flux increases with decreasing line width, which is consistent with the narrow-line objects being more face-on. This provides strong support for the equatorial scattering model. \citet{Afanasiev+Popovic15} point out the independence of the width of broad lines in polarized flux removes the uncertainty in black hole mass estimates due to the unknown inclination of the BLR.

\citet{Smith+02,Smith+05} and others assumed that the scattering ring was perfectly smooth and axisymmetric (see Fig.~25 in \citealt{Smith+02} and Fig.~1 in \citealt{Smith+05}).  However, there is strong evidence that absorbing gas and dust close to the centers of AGNs is clumpy and hence non-axisymmetric (see \citealt{Gaskell+Harrington18}).  We point out that {\em the scattering regions will therefore also be clumpy.}  As a consequence, PA swings across broad-line profiles will not be exactly the symmetric sideways-S shaped swings shown by models. The PA swings can be asymmetric and hence slightly offset in velocity. For the purpose of finding directions of AGN rotation, we assume that equatorial-scattering, as we have just discussed, is the main cause of polarization swings across the broad Balmer lines, and that deviations from the predicted simple ``sideways S'' dependence are due to non-axisymmetry in the scattering dust.  Because some degreee of line-profile asymmetry in AGNs is common, we do not expect most AGNs to have as clear PA rotations as the example in Fig.~2.  The perturbations on the predictions of the simplest equatorial-scattering model could potentially be large enough to give an incorrect direction of rotation but inspection of the high-quality \citet{Afanasiev+19} sample shows that AGNs not showing the sideways S are rare. We only have a few ambiguous cases in our sample (see Section 6)

The commonly observed slow changes in broad Balmer line profiles on timescales of months and years imply that the dust clumps are moving \citep{Gaskell+Harrington18}.  This can result in related changes in the polarization.  Details of this are beyond the scope of the present paper and we defer discussion of this for future work. 

\section{Data samples}

Thermal AGNs \citep{Antonucci12} can be divided into type-1 and type-2 on the basis of the appearance of their optical spectra \citep{Khachikian+Weedman74}. Type-1 AGNs have broad Balmer line profiles, while type-2 AGNs have only narrow lines.  The polarization properties of type-1 and type-2 AGNs are quite different. Type-2 Seyferts have higher continuum polarization than type-1s, and the polarization position angles are generally perpendicular to the projected radio axis, whereas type-1s show lower polarization with angles roughly aligned with the radio structure \citep{Stockman+79,Antonucci83,Antonucci84}.  The polarization differences are due to differing viewing angles.  In type-1 AGNs we get a direct view of the inner regions and the polarization parallel to the axis of symmetry is a result of scattering in the equatorial plane \citep{Antonucci84}.  In type-2 AGNs the inner regions are obscured by the surrounding dust and the polarization is caused by scattering off the far side of the dust \citep{Antonucci83}. We are only interested in AGNs with a direct view of the BLR, so we excluded type-2 (and type-1.9) AGNs from our study.

We carried out a literature search to identify AGNs with suitable spectropolarimetry for determining directions of rotation and then searched for the highest-quality images to infer the direction of each AGN’s host galaxy rotation. 

\section{Determining directions of rotation}  

As described in Section 3, the direction of rotation of each AGN’s BLR was determined from the PA swing across H$\upalpha$ or H$\upbeta$.  For the best available spectropolarimetry we compared the average PA between the line center (zero velocity) and the maximum velocity of the blue wing to the average PA between the line center and the maximimum velocity of the red wing. If the red side had the highest average PA, the rotation is clockwise; if the blue side had the highest average PA, the rotation is counter-clockwise.  In only a few cases (see below) was the difference in PA uncertain.
 
The direction of rotation of the host galaxies can be determined in two ways. If spiral arms are visible, we infer direction of rotation by assuming the spiral arms are trailing. This can only be done for disk galaxies of course.  If spatially-resolved velocity information is available, and it is clear which side of the galaxy is nearest to us, this can also be used to determine the rotation of direction. In one case, we used spatially-resolved velocity information to determine host-galaxy rotation directions because we could not determine the direction from imaging alone. 

\section{Results}

The top half of Table~1 summarizes the inferred rotation properties of the AGNs in alphabetical order. We list the direction of rotation of the AGN inferred from spectropolarimetry, and of the host galaxy inferred from imaging, along with references to data sources. We also indicate whether the AGN and host galaxy are co-rotating or counter-rotating. Additional columns include black hole mass estimates and notes on individual objects. Objects for which the host galaxy rotation could not be reliably determined are listed separately in the bottom half of the table. For a few cases we put colons after inferred AGN rotation directions where the PA swings are less certain.  Because of the small number of cases where the direction of AGN rotation could not be determined, the omission of these AGNs has no effect on our results.

We also note two AGNs (NGC 4151 and Mrk 231) for which the spectropolarimetry from different epochs is in conflict (see discussion below). In these cases, measurements taken at different times suggest opposite or inconsistent polarization signatures, preventing a unique determination of the rotation sense.  Exclusion of these AGNs, or of the few AGNs where the determination of rotation directions is less certain (marked with colons in the table), has no effect on our results.

From Table 1 it can be seen that the spectropolarimetry gives 27 cases of clockwise rotation and 27 cases of counter-clockwise rotation.  For AGNs where we also have an inferred direction of host-galaxy rotation, we find 11 co-rotating AGNs and 13 counter-rotating ones. 


\begin{table*}[htbp]
\caption{Rotation directions}
\resizebox{\textwidth}{!}{
\begin{tabular}{@{}lclclcccll}
\hline	
Object name	& AGN rotation& Specpol.& Galaxy rotation & Image &  Co-rotate? & Contra? & $\log M_{bh}$& Mass & Notes	\\

\hline
Akn 120, Mrk 1095	&	CW	&	1,2,21	&	CCW	&	1	&		&	1	&	8.33	&	1	&		\\
Akn 564	&	CCW	&	1	&	CW	&	1	&		&	1	&	6.42	&	3	&		\\
Fairall 51	&	CW	&	1,3,4	&	CW	&	1	&	1	&		&	7.71	&	4	&	Polarization high	\\
IRAS 13349+2438	&	CW	&	2	&	CW:	&	3	&	1	&		&	9.10	&	26	&		\\
I Zw 1	&	CW	&	2	&	CW	&	1	&	1	&		&	7.46	&	1	&		\\
MCG+08-11-011	&	CCW	&	2	&	CW	&	2	&		&	1	&	8.19	&	1	&		\\
Mkn 79	&	CCW	&	2	&	CW	&	3	&		&	1	&	7.45	&	1	&		\\
Mkn 704	&	CCW	&	2,6,21	&	CCW	&	1	&	1	&		&	7.63	&	5	&		\\
Mkn 817	&	CCW	&	2	&	CW	&	3	&		&	1	&	7.65	&	1	&		\\
Mrk 985	&	CW	&	1	&	CCW	&	4	&		&	1	&	7.65	&	4	&		\\
Mrk 1044	&	CCW	&	5	&	CCW	&	1	&	1	&		&	6.28	&	4	&		\\
Mrk 1048	&	CW	&	21	&	CCW	&	3	&		&	1	&	7.66	&	23	&		\\
Mkn 1501	&	CW	&	2	&	CW	&	2	&	1	&		&	7.86	&	24	&		\\
NGC 1097	&	CCW	&	8,19	&	CW	&	1	&		&	1	&	8.15	&	6	&		\\
NGC 3227	&	CW	&	2,3	&	CW	&	1	&	1	&		&	7.31	&	1	&		\\
NGC 3516	&	CW	&	21	&	CCW	&	6, 8	&		&	1	&	7.40	&	21	&	Ambiguous image; used velocity field	\\
NGC 4051	&	CCW	&	2	&	CCW	&	2	&	1	&		&	6.24	&	10	&		\\
NGC 4593	&	CCW	&	1,2	&	CW	&	1	&		&	1	&	7.28	&	1	&		\\
NGC 5548	&	CCW	&	1,2,6,21	&	CCW	&	1	&	1	&		&	7.81	&	1	&		\\
NGC 7469	&	CCW	&	1	&	CW	&	1	&		&	1	&	7.52	&	2	&		\\
PG 0844+349	&	CW	&	2	&	CW	&	3	&	1	&		&	7.96	&	10	&		\\
VII Zw 244	&	CW	&	8	&	CCW	&	2	&		&	1	&	8.29	&	7	&		\\
Was 45 (UGC 7064)	&	CW	&	3	&	CCW	&	1	&		&	1	&	7.66	&	25	&		\\
3C 120	&	CCW	&	2	&	CCW	&	1	&	1	&		&	7.68	&	1	&		\\

\hline
Arp 102B	&	CCW	&	10	&		&		&		&		&	8	&	17	&		\\
ESO 242-G35	&	CCW	&	1	&		&		&		&		&		&		&		\\
Fairall 9	&	CW	&	12	&		&		&		&		&	8.41	&	14	&		\\
IRAS 03450+0055	&	CCW	&	2	&		&		&		&		&		&		&		\\
IRAS 04416	&	CW	&	5	&		&		&		&		&	6.97	&	21	&		\\
KUV 18217+6419	&	CW	&	1	&		&		&		&		&		&		&		\\
LEDA3095839	&	CW:	&	8	&		&		&		&		&		&		&		\\
Mrk 6	&	CCW	&	1,2,14,15	&		&		&		&		&		&		&		\\
Mkn 110	&	CCW	&	2	&		&		&		&		&	7.26	&	12	&		\\
Mrk 231	&		&	2,3,6,7,21	&	CCW	&	2	&		&		&		&		&	Conflicting specpol   \\
Mrk 279	&	CCW	&	1	&		&		&		&		&	7.58	&	15	&		\\
Mrk 304	&	CW	&	1,2,21	&		&		&		&		&		&		&		\\
Mrk 335	&	CCW	&	1,2	&		&		&		&		&	7.25	&	13	&		\\
Mrk 376	&	CW	&	6,21	&		&		&		&		&		&		&		\\
Mrk 486	&	CW	&	21	&		&		&		&		&		&		&		\\
Mrk 509	&	CW	&	1,2,6,21	&		&		&		&		&	8.04	&	10	&		\\
Mkn 668	&	CCW	&	2	&		&		&		&		&		&		&		\\
Mrk 841	&	CCW	&	1,2	&		&		&		&		&	7.67	&	3	&		\\
Mrk 876	&	CCW	&	1,2	&		&		&		&		&		&		&		\\
Mkn 1148	&	CW	&	2	&		&		&		&		&		&		&		\\
Mrk 1239	&	CW:	&	17,18	&		&		&		&		&		&		&	Unclear rotation	\\
MS 1849.2-7832	&	CW	&	1	&		&		&		&		&		&		&		\\
NGC 3783	&		&	1	&	CCW	&	1	&		&		&	7.47	&	2	&		\\
NGC 4151	&		&	2,21	&	CW	&	4	&		&		&	7.08	&	1	&	Conflicting specpol	\\
PG 1700+518	&	CCW	&	2, 23	&		&		&		&		&		&		&		\\
PKS 2004-447	&	CW	&	20	&		&		&		&		&		&		&		\\
SDSS J080101.41+184840.7	&	CCW	&	5	&		&		&		&		&	6.4	&	20	&		\\
3C 273	&	CCW	&	2	&		&		&		&		&	9.82	&	8	&		\\
3C 332	&	CW	&	10	&		&		&		&		&	8.64	&	23	&		\\
3C 351	&	CW	&	10	&		&		&		&		&		&		&		\\
3C 390.3	&	CCW	&	10,11	&		&		&		&		&	9.23	&	9	&		\\
3C 445	&	CCW	&	22, 2	&		&		&		&		&		&		&		\\
4C 73.18	&	CW	&	10	&		&		&		&		&	8.57	&	16	&		\\
\hline					
\end{tabular}
}
\end{table*}


\clearpage
\onecolumngrid

\bigskip
\noindent{\bf Table 1. Reference Codes}

\medskip
{\bf Spectropolarimetry:}

(1) \citealt{Smith+02};
(2) \citealt{Afanasiev+19};
(3) \citealt{Smith+04};
(4) \citealt{Schmid+01};
(5) \citealt{Sniegowska+23};
(6) \citealt{Goodrich+Miller_94};
(7) \citealt{Smith+95};
(8) \citealt{Shablovinskaya+22};
(9) \citealt{Young+96};
(10) \citealt{Corbett+00};
(11) \citealt{Afanasiev+15};
(12) \citealt{Jiang+21};
(13) \citealt{Pan+19};
(14) \citealt{Smith+05};
(15) \citealt{Afanasiev+14};
(17) \citealt{Goodrich89};
(18) \citealt{Robinson+11};
(19) \citealt{Young+96};
(20) \citealt{Baldi+16};
(21) \citealt{Martel96};
(22) \citealt{Cohen+99};
(23) \citealt{Young+07}.

\medskip
{\bf Images:}

(1) DESI Legacy Imaging Surveys DR10 (via Aladin Lite);
(2) Pan-STARRS DR1 (via Aladin Lite);
(3) SDSS DR9 (via Aladin Lite);
(4) DSS2 Blue (via Aladin Lite);
(5) GALEX GR6/7 (via Aladin Lite);
(6) \citealt{WikiNGC3516};
(7) \citealt{Esposito+24};
(8) \citealt{Mulchaey+92}.

Host-galaxy images were accessed through the Aladin Lite interface \citep{AladinLite}. 
The imaging data listed above correspond to surveys made available within Aladin.

\medskip
{\bf Masses:}

(1) \citealt{Afanasiev+19};
(2) \citealt{Vestergaard+Peterson06};
(3) \citealt{Piotrovich+15};
(4) \citealt{Bennert+06};
(5) \citealt{DeRosa+18};
(6) \citealt{Onishi+15};
(7) \citealt{Shablovinskaya+22};
(8) \citealt{Paltani+05};
(9) \citealt{Sergeev+17};
(10) \citealt{Peterson+04};
(11) \citealt{Denney+09};
(12) \citealt{Kollatschny04};
(13) \citealt{Grier+12};
(14) \citealt{Koratkar+89};
(15) \citealt{Runnoe+13};
(16) \citealt{Shapovalova+13};
(17) \citealt{Zhang+11};
(18) \citealt{Du+15};
(19) \citealt{Sniegowska+23};
(20) \citealt{Halpern90};
(21) \citealt{Bentz+Katz15};
(22) \citealt{Pandey+25};
(23) \citealt{Grier+17};
(24) \citealt{LaMassa+25};
(25) \citealt{Lee+13};
(26) \citealt{Kawakatu+07}.

\medskip

{\bf Notes:} (1) Irregular profile with one or two displaced humps.

\clearpage
\twocolumngrid

\section{Discussion}

\subsection{NLR Outflows}
The fact that outflowing NLR bicones also show a ``zone of avoidance'' (see Section 2), as do the radio jets, implies that whatever is the cause of this is not something uniquely connected to radio jets.

\subsection{The cause of PA rotations across broad lines}

The similarity we find for the numbers of greatest PAs for the blue and red sides of the lines is important because it strongly supports the equatorial-scattering model.  Swings in PA over a single limited velocity range across a line were initially interpreted as scattering off bulk inflows or outflows (i.e., by \citealt{Martel98}), but the equal numbers of increases and decreases in PA on a given side of the line makes such interpretations highly unlikely. 

\subsection{Black hole rotations}
 
As explained in the Introduction, if the gas feeding the AGN preserves the host galaxy’s angular momentum, co-rotation should dominate; while if the angular momentum vector direction is randomized, co-rotation and counter-rotation should occur in equal numbers. The nearly equal numbers of co-rotating and counter-rotating AGNs we find thus strongly favors the latter scenario, indicating that the orientations of the AGNs we consider are effectively decoupled from those of the host galaxy. Spectropolarimetry provides a more stringent test of alignment than axis comparisons alone and confirms that AGN rotation directions are effectively random with respect to their host galaxies, at least for our sample.

The lack of alignment is inconsistent with simple, coherent accretion scenarios, which would predict that angular momentum is largely preserved from galactic to nuclear scales. Instead, our results suggest that accretion is frequently chaotic, with gas arriving in discrete, randomly-oriented episodes, and with there being black-hole mergers. This interpretation directly influences theoretical models: it implies that black hole spin growth is not dominated by long-term, aligned inflow but by irregular accretion events and mergers, which affects predictions for radiative efficiency, feedback, and hence galaxy evolution.

\subsection{Mass dependence?}

Theoretical work (see Introduction) predicts that the relative orientation of spins will depend on the history of a galaxy and as a result could be a function of black hole mass.  For example, low mass spirals with low-mass black holes could have better alignment than higher mass black holes and galaxies.  We have therefore given black hole masses in Table 1.  Unfortunately, the range of black hole masses of AGNs with existing spectropolarimetry is limited and the sample is too small to see if there is an effect.

We suggest that investigating mass dependence could be a useful area for further study.  For nearby, low-mass spiral galaxies there is no difficulty in determining the directions of galaxy rotation and obtaining more spectropolarimetry is all that is needed.

\subsection{Nature of the scattering regions}

The simple equatorial scattering model predicts smooth, symmetric, ``sideways-S'' profiles, which are rarely observed in practice. We propose that rather than being due to additional outflowing or inflowing components, the frequent departures from the simple prediction are indicating instead that the {\em scattering regions are generally clumpy and not axisymmetric.} \citet{Gaskell+Harrington18} have shown that slow changes in broad-line profiles, velocity-dependant Balmer decrements and in velocity-dependent time delays can all be explained by the motion of clumps of dusty material close to the BLR. The existence of such moving clumps is supported by the well-known variability of X-ray absorption (see references in \citealt{Gaskell+Harrington18}).  On average, though, the scattering from this clumpy dust will resemble scattering from a smooth, axially-symmetric equatorial scatter.

\subsection{Spectropolarimetric variability}

The moving-clump model obviously predicts that PA swings will vary with time. Slow changes in broad-line profiles, in velocity-dependant Balmer decrements, and velocity-dependent time delays all occur on timescales of months to years (see \citealt{Gaskell+Harrington18}). We therefore predict that the details of PA swings will vary on similar timescales. \citet{Martel98} has discussed such possible changes in NGC~4151 in the 1990s. The observations of \citet{Afanasiev+19} two decades later, suggest a larger change. Another example of probable  variability is provided by the spectropolarimetry of Mrk 231 (see references in Table 1). However, we caution that polarization PA is determined from the ratio of the $Q$ and $U$ Stokes parameters, both of which are small numbers. Small errors due to observational effects (such as changes in seeing and objects centering in the spectrograph slit), difference in processing and noise can all produce spurious changes in PA. In the cases of NGC~4151 and Mrk~231 the apparent changes in PA are large enough to give uncertainty in the direction of rotation of the AGN.  Careful spectropolarimetric monitoring on the expected timescale would help in testing the moving-clump model and the off-axis flare model (see \citealt{Gaskell10} and \citealt{Goosmann+14})

\section{Conclusions}

We have used spectropolarimetry to determine the directions of rotation of AGNs and compared them with the inferred directions of rotation of their host galaxies. 

For the full sample, we find 27 AGNS rotating clockwise and 27 counter-clockwise. For the subsample where host galaxy rotation could also determined, we find 11 co-rotating and 13 counter-rotating systems. These similar numbers show that AGN rotation directions are not systematically related to those of their host galaxies.

This result rules out the idea that the observed ``zone of avoidance'' for radio jets (which we also show is present for NLR outflows) is due to AGN spin axes generally remaining close to the host-galaxy rotation axis. Instead, it supports a picture in which the orientations of AGNs are effectively random with respect to their hosts. 

Our results support the equatorial-scattering interpretation of polarization PA swings across broad lines. We propose that departures from a simple symmetric ``sideways-S'' pattern are a result of the scattering regions being somewhat clumpy and not perfectly axisymmetric. Possible multi-epoch changes in the PA behavior of objects such as NGC~4151 and Mrk~231 are consistent with this picture and motivate future spectropolarimetric monitoring.

\begin{acknowledgments} 

We wish to thank Ski Antonucci, Julie Biedermann, Sandra Faber, Fr\'ed\'eric Marin, Luka Popovi\'{c}, the late Joel Primack, and Elena Shablovinskaya for helpful comments and feedback.

\end{acknowledgments}

\clearpage
\FloatBarrier
\bibliography{Rotation_01}

\end{document}